\documentclass[fleqn,usenatbib]{mnras}

\usepackage{newtxtext,newtxmath}

\usepackage[T1]{fontenc}

\DeclareRobustCommand{\VAN}[3]{#2}
\let\VANthebibliography\thebibliography
\def\thebibliography{\DeclareRobustCommand{\VAN}[3]{##3}\VANthebibliography}

\usepackage{graphicx}	
\usepackage{amsmath}	
\usepackage{pgfplots}	
\usepackage{xcolor}
\usepackage{breakurl} 

\pgfplotsset{compat=1.16}

\title[Optical Variability of 1H0323+342]{Optical Variability of the very Radio-Loud Narrow line Seyfert 1 galaxy, 1H 0323+342}

\author[C. S. Turner et al.]{
Clay S. Turner,$^{1}$\thanks{E-mail: turner@astro.gsu.edu (CST)}
Hugh R. Miller,$^{1}$
Jeremy D. Maune$^{1}$
and Joseph R. Eggen$^{2,3,4}$
\\
$^{1}$Georgia State University, Department of Physics and Astronomy, Atlanta, Georgia 30303, USA\\
$^{2}$Department of Astronomy, University of Maryland, College Park, MD 20742, USA\\
$^{3}$Center for Research and Exploration in Space Science and Technology, NASA/GSFC, Greenbelt, MD 20771, USA\\
$^{4}$NASA Goddard Space Flight Center, Greenbelt, MD 20771, USA\\
}

\date{Accepted XXX. Received YYY; in original form ZZZ}

\pubyear{2022}

\begin{document}
\label{firstpage}
\pagerange{\pageref{firstpage}--\pageref{lastpage}}
\maketitle

\begin{abstract}
1H 0323+342 is optically one of the nearest and brightest very radio
loud narrow line Seyfert 1 galaxies (vRL NLSy1). It is also one of the first vRL
NLSy1s detected at gamma-ray energies by the Fermi-LAT. We report the results
of monitoring the optical flux of 1H 0323+342 during more than six and a half
years. In some cases, we, for the first time, simultaneously use two telescopes to
monitor the optical flux of 1H 0323+342 on timescales ranging from minutes to
hours, demonstrating the reality of low amplitude microvariability whole events
with durations of a few hours for this object.  Based on the present results, as well as those of
earlier studies, we suggest that this represents a fundamental timescale associated with
the underlying source region We also present an enhancement of
Howell’s comparison star method for detecting microvariability.
\end{abstract}




\section{Introduction}

The very radio loud narrow line Seyfert 1 galaxy (vRL NLSy1) 1H 0323+342 is one of the 
first of this class to be detected at gamma-ray energies by the Fermi-LAT \citep{Abdo2009}.  
Previously, gamma ray loud active galactic nuclei (AGN) have been identified most often with BL Lac 
objects and flat spectrum radio quasars, i.e. blazars.   However, recently a small number of radio 
galaxies have also been detected with Fermi. The host galaxies for all these classes are elliptical galaxies. 
The host galaxies associated with most Seyfert galaxies are typically spiral galaxies. Since gamma ray 
emission is thought to be the signature of the presence of a relativistic jet oriented near the 
line-of-sight to the observer, this indicates, for the first time, that spiral galaxies may also 
host these highly energetic jets.   vRLNLSy1s are a class of AGN with unusual characteristics when 
compared with those associated with Seyfert 1 galaxies \citep{Komossa2018,Pogge2000}. 
The host galaxies for NLSy1 galaxies are usually spiral or barred spirals 
\citep{Ferrara2000,Crenshaw2004,Deo2006}. The $H\beta$ emission line has
FWHM<2000 $km s^{-1}$ \citep{Goodrich1989} and their spectra contain strong Fe II emission \citep{Osterbrock1985}. 
They have steep soft X-ray spectra, and exhibit strong, large
amplitude X-ray variability, (e.g., \citealt{Boller1996} ).
These extreme properties can
be understood in terms of vRL NLSy1s having relatively low-mass super massive
black holes (SMBH) and accretion rates near the Eddington accretion rate \citep{Komossa2008}.

\citet{Yuan2008} defined vRL NLSy1 galaxies as those NLSy1 galaxies
which have R>100, where R is defined as the ratio of rest frame flux at 1.4 GHz
to the optical flux in the B band at 4400\AA . 
Thus, those members of vRL NLSy1s detected at gamma ray 
energies share many properties similar to those seen for blazars
\citep{Abdo2009,Foschini2015}. In particular, blazars exhibit large amplitude,
rapid variations with timescales as short as minutes-to-hours. 
Rapid and large amplitude optical variability is a well-established property for blazars and
blazar-like AGN such as vRL NLSy1s \citep{Smith1996}. Indeed, rapid optical variations
on timescales as short as minutes to hours, known as microvariability \citep{Miller1989} 
or intranight optical variability (INOV) \citep{Wagner1995}, have been
reported for several vRL NLSy1 galaxies including 1H 0323+342, e.g. \citet{Ojha2019,Paliya2013,Paliya2014,Maune2014}. 
1H 0323+342 is one of the nearest vRL NLSy1 galaxies detected at gamma ray energies and is among the 
brightest of this class observed at optical wavelengths \citet{Foschini2015,Doi2020}.
As such, it is an attractive object to monitor for rapid variations at optical wavelengths. Variability studies, such as those reported in the present study, provide valuable insight into the physical mechanism(s) responsible for the emission processes and the structure and size of the emitting region(s).  
However, none
of the earlier investigations demonstrated the reality of well-defined whole
events (i.e., variations which exhibit both the rise and decline in flux defining the
event) with optical ranges as small as $\approx0.03$ mag and a duration as short as a few
hours such as are presented in this present investigation. 

In this paper, we report the results of a monitoring program begun in 2010,
during which we have monitored the optical flux of 1H0323+342 on timescales as
short as minutes to as long as years. On one night we obtained a unique set of
observations using multiple telescopes to provide quasi-simultaneous monitoring
of the optical flux on ~minute timescales. Previously \citet{Miller2017a} presented
preliminary results using a small, early segment of the data. The present paper
builds upon those earlier reported results and includes a substantial body of new
unpublished observations.

\section{Optical Observations and Data Reduction}

We obtained the majority of the photometric observations included here using the 31-inch NURO telescope, 42-inch Hall telescope, and the 72-inch Perkins telescope at Lowell Observatory in Flagstaff, AZ.  We collected additional optical data using the 24-inch Plane Wave Instruments CDK telescope located at Georgia State University’s Hard Labor Creek Observatory (HLCO) near Rutledge, GA. The Perkins, Hall, and NURO telescopes are equipped with custom designed cryogenically chilled CCD cameras fitted with Johnson filters for photometry. Similarly, the 24-inch Planewave telescope uses a commercially available Peltier chilled CCD camera (Apogee Alta U230) with Johnson filters.

The observational data presented here for 1H 0323+342 consists of nearly 5000 individual observations obtained on 85 different nights from November, 2010 to March, 2017. Nightly summary data is given in Table \ref{tab:NightlySummary}. In the table we provide mean Julian date, the common date, number of nightly observations,  time span in minutes for the observing run, mean R band magnitude, and, for nights with five or more observations, the modulation index percentage defined in \citet{Quirrenbach2000}.  These observations were included as part of our larger blazar monitoring program which scheduled approximately one week of observations, less than a week away from the new moon (dark time), each month.    The observations consist of nights when as few as two observations per night were obtained to nights when hundreds of images were taken at minute-scale time resolution for a duration of several hours during that night. On several occasions, quasi-simultaneous observations were also taken with two different telescopes in an effort to provide independent confirmation for the reality of any small amplitude (<~0.03 mag) microvariability events which might have timescales from minutes to hours.

The R-band magnitudes were derived using differential aperture photometry with in-field comparison stars. We selected the comparison stars based on each of them having a brightness comparable to 1H 0323+342 and for their exhibiting photometric stability over long periods of time. 
The R-band magnitudes for the comparison stars were calibrated using \citet{Landolt1992} standard stars. The finding chart for 1H 0323+342 along with the comparison stars which we used and their R-band magnitudes may be found on our group’s dedicated web pages at \url{https://sites.google.com/site/jdmaune/seyfert-fields/0323-342}.  At the beginning of each night, we collected a set of 10 bias frames and subsequently combined them into a single super bias frame.  In addition, we also collected a set of 10 flat field images during the night and combined them to form a single super flat field frame.  The super bias frame was then subtracted from each individual object frame and was then divided by the super flat field frame.  We utilized IDL for most of the data reduction.  However, we used MaxIm DL and IRAF for a small portion of the data. Maxim DL and IRAF have been shown to produce nearly identical results \citep{Blackwell2005}.  Similarly, we have reduced some data using IDL and Maxim DL (specifically for the nights of simultaneous observations), compared the results, and found there was good agreement in the results for IDL and Maxim DL. The similar results for the various programs is not surprising since differential aperture photometry is a basic form of analysis. For consistency, we standardized to a seven arcsecond diameter aperture for the analyses of all the data taken with any of the four telescopes.

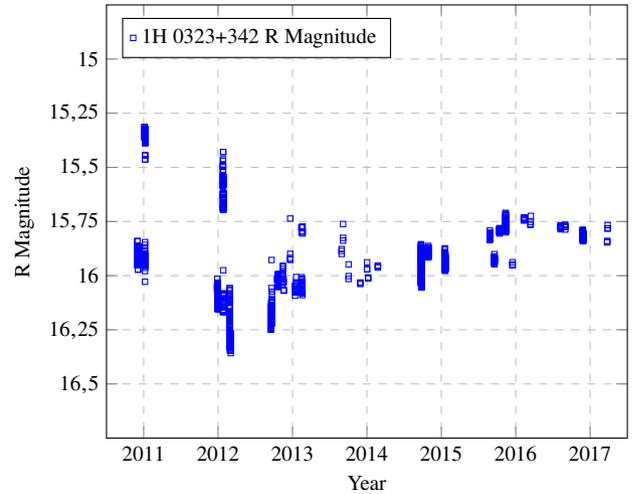
\begin{figure}
\begin{tikzpicture}
\begin{axis}[  
    xlabel={Year},
    ylabel={R Magnitude},
    xmin=2010.5, xmax=2017.5,
    ymin=14.75, ymax=16.75,
    xtick={2011,2012,2013,2014,2015,2016,2017,2018},
    ytick={15,15.25,15.5,15.75,16,16.25,16.5},
    legend pos=north west,
    ymajorgrids=true,
    xmajorgrids=true,
    grid style=dashed,
    y dir=reverse,
	/pgf/number format/.cd,
    	    use comma,
        	1000 sep={}    
]
\addplot[
	only marks,
    color=blue,
    mark=square,
    mark size=1.0pt
    ]
    table {0323+342_TotalSetDate.prn};
    \legend{1H 0323+342 R Magnitude}
    
\end{axis}
\end{tikzpicture}
\caption{1H 0323+342 R band light curve}
\label{fig:Master}
\end{figure}

\section{R Band Light Curve}

The range for the optical variability for 1H 0323+342, covering an interval of $\approx 6
$ years, is shown in  Fig. \ref{fig:Master}.  This source is seen to be extraordinarily active from late 2010 to early 2013 with several rapid, large amplitude events observed where the source varies $\approx 1.0$ mag on the timescale of less than a week.   However, the variations appear to be of lower amplitude and tend to occur when the object is near the fainter end of the range exhibited in Fig. \ref{fig:Master} after 2013.  Although this seems to be true for the overall variability patterns shown in figure \ref{fig:Master}, there are exceptions to this general trend.  For example, 1H 0323+342 appears to exhibit extremely rapid large-amplitude variations on Feb 19, 2012 while it is in a bright state and similar, large-amplitude variations are observed on Feb 21, 2012 when the source was near a minimum of its brightness.  The source seems to be unusually active during this period.  Unfortunately, the monitoring of 1H 0323+342 is not sufficiently dense and continuous to determine if there is any relationship between the brightness/faintness of the object and the presence of rapid large amplitude variations. 

\section{Test for the Presence of Microvariability}

\begin{figure}

\begin{tikzpicture}
\begin{axis}[
    xlabel={Hours Post Common Start Time (JD2456923.797)},
    ylabel={R Magnitude},
    xmin=0, xmax=4,
    ymin=15.8, ymax=16.0,
    xtick={0,1,2,3,4},
    ytick={15.8,15.85,15.9,15.95,16.0},
    legend pos=north west,
    ymajorgrids=true,
    xmajorgrids=true,
    grid style=dashed,
    y dir=reverse,
]
\addplot[
	only marks,
    color=blue,
    mark=square,
    mark size=1.0pt
    ]
    table {T31_hourspostcommon.prn};
    \addlegendentry{31" Telescope}

\addplot[
	only marks,
    color=red,
    mark=square,
    mark size=1.0pt
    ]
    table {T72_hourspostcommon.prn};
    \addlegendentry{72" Telescope}
    
\end{axis}
\end{tikzpicture}

\caption{31" and 72" R band magnitudes during flare}
\label{fig:FlareMag}
\end{figure}
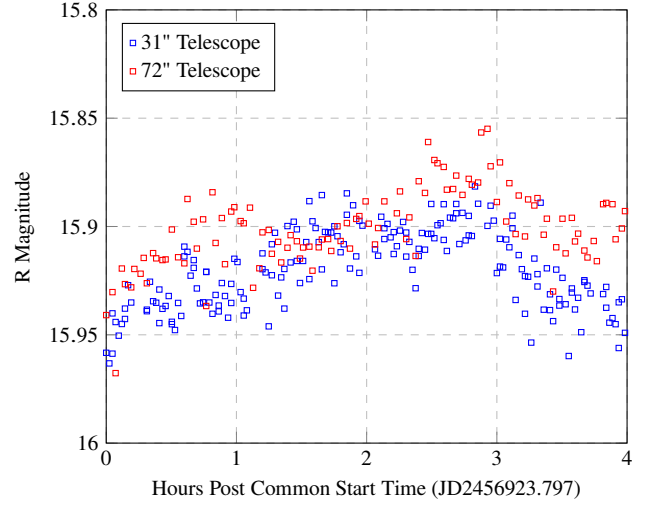

A visual inspection of the light curve shown in Fig. \ref{fig:Master} clearly indicates that 1H 0323+342 exhibits variability over timescales ranging from less than a day to years. Several earlier variability studies of 1H 0323+342, e.g., \citep{Paliya2013,Ojha2019}, seeking to identify microvariations, have found positive evidence for the existence of small amplitude
 ($\approx0.1$ mag) variations on the timescale of minutes to hours. We propose to investigate the question of whether small, lower amplitude flares with durations of only a few hours or less can be identified with events that are statistically significant and therefore real. We designed a set of observations using two geographically separated telescopes, where each telescope observed 1H 0323+342 during the same time on the same night. Specifically, the telescopes we used were the 72-inch and the 31-inch telescopes located at Lowell Observatory. The telescopes are located just over 250 meters apart on Anderson Mesa, AZ. On 23 Sept 2014, we monitored 1H 0323+342 with both telescopes for an interval of just over 4 hours. During the time 1H 0323+342 was monitored, a small, low-amplitude flare was observed. Figure \ref{fig:FlareMag} shows the R band magnitude vs the time in hours measured with each telescope. The time scale on the x-axis is measured relative to the common start time. Upon inspection of Fig. \ref{fig:FlareMag}, one sees that the 31-inch telescope data and the 72-inch data are sampled asynchronously with respect to each other. The 31-inch telescope’s data set consists of a total of 297 observations and the 72-inch telescope’s data set contains a total of 115 observations.
This is due to each telescope using different exposure times and different sampling rates. However the flare present in each data set is visually discernible nonetheless. In order to determine the statistical significance of the variations, we have analyzed this event in two different ways. First we wish to see if the observed variance during the flare for the AGN is larger than the expected variance a reference star would have if its magnitude were the same as the AGN’s mean magnitude during this period. 

\begin{figure}
\begin{tikzpicture}
\begin{semilogyaxis}[
    title={72" Data Variances},
    xlabel={Mean R Magnitude},
    ylabel={Variance},
    xmin=14.5, xmax=17.5,
    ymin=1.0E-6, ymax=2.0E-4,
    xtick={14,14.5,15,15.5,16,16.5,17,17.5},
    ytick={},
    legend pos= south west,
    ymajorgrids=true,
    xmajorgrids=true,
    grid style=dashed,
    x dir=reverse,
]

\addplot[
	domain=14.5:17.5,
	samples=100,
	color=green,
	]
	{10^(-15.993+0.689*x)};
   \addlegendentry{Regression Line}

\addplot[
	only marks,
    color=blue,
    mark=square,
    mark size=1.0pt
    ]
    coordinates {
   
	(16.418,1.838E-5)
	(15.826,1.06E-5)
	(14.76,1.409E-6)
	(17.325,9.061E-5)
	(16.237,1.49E-5)
	};

    \addlegendentry{Comparison Stars}
    
\addplot[  
	only marks,
    color=black,
    mark=square,
    mark size=2.0pt
    ]
    coordinates { 
	 (15.739,1.558E-4)    
    };
    \addlegendentry{AGN}
     
\end{semilogyaxis}
\end{tikzpicture}
\caption{72-inch telescope variance vs R band magnitude}
\label{fig:Variance72}
\end{figure}
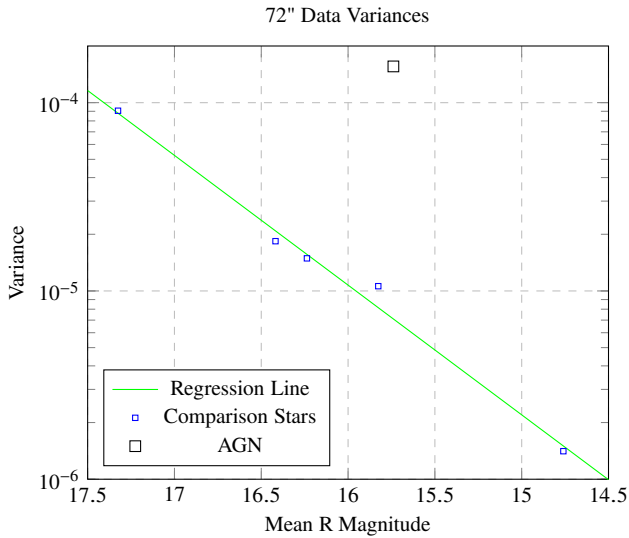

Following \citet{Howell1986}, we obtained a series of images with each telescope. Each image is reduced, as described above, with the magnitude data being placed into vectors based upon which reference star, AGN and which telescope was used. E.g., the 31" telescope data during the flare consists of 297 measurements for each of the reference stars and for the AGN. Thus the 297 values for Reference star "A" are used to create a vector. Likewise the 297 values for reference star "B" are used to make a vector. This is done in turn for each reference star and for the AGN. Then a mean magnitude and variance is calculated per vector. The 72" data is treated similarly.
\citet{Howell1988}, shows the variances of the reference stars' magnitudes scale with
their magnitudes. I.e., the brighter the star, the lower the variance will be for a given exposure time. Thus, we can fit a power law to the set of magnitudes and variances and use it to predict the variance of a reference star as if its magnitude were the same as the AGN’s mean magnitude. 
The results of the power law fits and the variances are shown in Figs. \ref{fig:Variance31} and \ref{fig:Variance72}. The numerical values for the predicted and measured variances are in Table \ref{tab:Varstat}. We can use an F-test \citep{Johnson1995} to compare the AGN’s observed variance and the predicted reference star’s variance to see if the observed AGN’s variability during the flare is likely due to the variances observed with the reference stars’ data or if the AGN has its own variance over and above that of the reference stars’ measured variances. Using a 99.5 per cent confidence interval 
(right hand $\alpha=0.005$) and our particular sample sizes we found the critical thresholds for the F scores and calculate the F scores. These are shown in Table \ref{tab:Varstat}. For both sets of data, the AGN’s variance is significantly larger than the critical threshold, thus we are confident that the object’s brightness is actually changing during this time frame.

\begin{table}
\caption{F-test and variance results}
\label{tab:Varstat}
\begin{tabular}{cccccc}
\hline\hline
 {\em Scope} & {$N$} & {$F$} & {$F_{\alpha=0.005}$} & Predicted Var. & Measured Var. \\ \hline
 31" & 297 & 4.675 & 1.35 & $8.011\times 10^{-5}$ & $3.745\times 10^{-4}$ \\
 72" & 115 & 21.717 & 1.62 & $7.173\times 10^{-6}$ & $1.558\times 10^{-4}$ \\ \hline\hline 
\end{tabular}
\end{table}

\begin{figure}
\begin{tikzpicture}
\begin{semilogyaxis}[
    title={31" Data Variances},
    xlabel={Mean R Magnitude},
    ylabel={Variance},
    xmin=14, xmax=17.5,
    ymin=3.0E-6, ymax=1.0E-3,
    xtick={14,14.5,15,15.5,16,16.5,17,17.5},
    ytick={},
    legend pos= south west,
    ymajorgrids=true,
    xmajorgrids=true,
    grid style=dashed,
    x dir=reverse,
]

\addplot[
	domain=14:17.5,
	samples=100,
	color=green,
	]
	{10^(-13.408+0.584*x)};
   \addlegendentry{Regression Line}

\addplot[
	only marks,
    color=blue,
    mark=square,
    mark size=1.0pt
    ]
    coordinates {
   
    (14.089,7.634E-6)
    (16.428,1.582E-4)
    (15.835,5.745E-5)
    (14.747,1.500E-5)
    (17.351,6.012E-4)
    (16.263,1.154E-4)
    };
    \addlegendentry{Comparison Stars}
    
\addplot[  
	only marks,
    color=black,
    mark=square,
    mark size=2.0pt
    ]
    coordinates { 
	 (15.946,3.745E-4)    
    };
    \addlegendentry{AGN}
     
\end{semilogyaxis}
\end{tikzpicture}

\caption{31-inch telescope variance vs R band magnitude}
\label{fig:Variance31}
\end{figure}
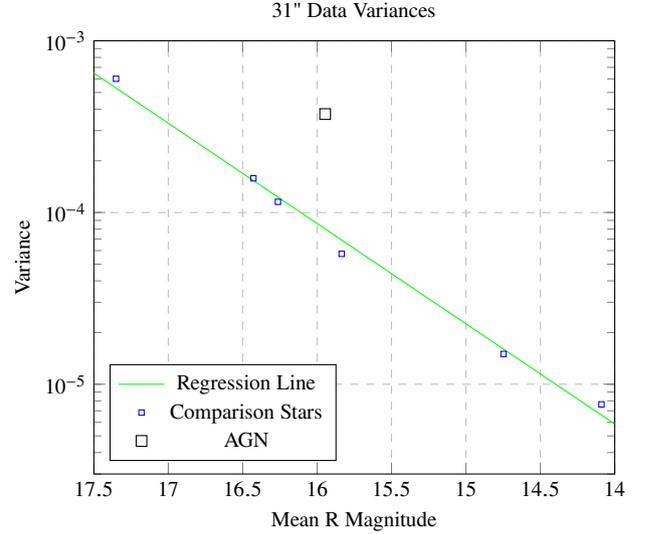

Secondly, we recognize that the intrinsic variability associated with 1H 0323+342 should be the same when detected at either telescope and therefore the light curves observed at each telescope should be highly correlated if there are intrinsic variations present. However the signal associated with the noise detected in the light curves produced at each telescope should be uncorrelated and not contribute to any correlation between the two light curves. In order to determine if a correlation is present, a discrete cross correlation test is used to find if both data sets varied together. We did this with the Z transformed discrete correlation function (ZDCF), \citep{Alexander1997}, which is an enhancement of \cite{Edelson1988}. The extension includes population sampling and Fisher’s z transformation of the Pearson coefficient. The DCF peaked at a time lag of zero minutes with a correlation coefficient (Pearson’s r value) of 0.806 and using Fisher’s z transformation we find the confidence interval of one standard deviation to be [0.767, 0.838]. This supports the reality of the 1H 0323+342 variability which is present in the data recorded from both telescopes. See fig. \ref{fig:ZTDCF} for the DCF with time lags up to $\pm60$  minutes.

The reader may notice that the mean magnitudes of 1H 0323+342 indicated in Figures \ref{fig:Variance31} and \ref{fig:Variance72} differ. This was caused by targeting an extended object and using too short exposures on the 31" telescope. Thus parts of the galaxy's light falls below threshold of measure resulting in the 31" telescope data for this run to be low. So we end up with an effective offset of under 0.2 magnitudes between the two data sets. Even with this difference, correlation of the two data sets is unaffected by such an offset.

\begin{figure}

\begin{tikzpicture}
\begin{axis}[
    xlabel={Relative Time Difference (minutes)},
    ylabel={Z Transformed Discrete Cross Correlation},
    xmin=-60, xmax=60,
    ymin=-0.5, ymax=1,
    xtick={-60,-45,-30,-15,0,15,30,45,60},
    ytick={-0.5,-0.25,0,.25,.5,.75,1},
    legend pos=south west,
    ymajorgrids=true,
    xmajorgrids=true,
    grid style=dashed,
]
\addplot[
	only marks,
    color=blue,
    mark=square,
    mark size=1.0pt
    ]
    table {DCF};
    \legend{Z Transformed Discrete Cross Correlation}
    
\end{axis}
\end{tikzpicture}

\caption{ZTDCT vs time lag ($\pm$ 60 minutes)}
\label{fig:ZTDCF}
\end{figure}
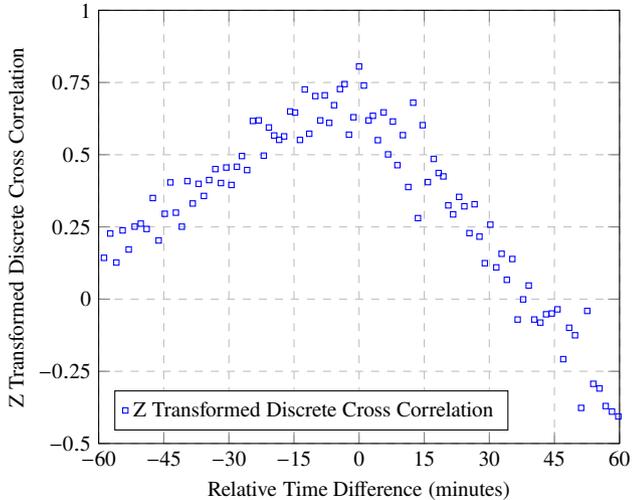

\begin{figure}
\begin{tikzpicture}
\begin{axis}[  
    xlabel={Year},
    ylabel={Modulation Index $m[\%]$},
    xmin=2010.5, xmax=2017.0,
    ymin=-1, ymax=25,
    xtick={2011,2012,2013,2014,2015,2016,2017,2018},
    ytick={0,5,10,15,20,25},
    legend pos=north east,
    ymajorgrids=true,
    xmajorgrids=true,
    grid style=dashed,
	/pgf/number format/.cd,
    	    use comma,
        	1000 sep={}    
]
\addplot[
	only marks,
    color=blue,
    mark=square,
    mark size=1.0pt
    ]
    table {0323+342ModIndexDate.prn};
    \legend{1H 0323+342 Modulation Indices $m[\%]$}    
\end{axis}
\end{tikzpicture}
\caption{1H 0323+342 Modulation Indices}
\label{fig:MasterModulation}
\end{figure}
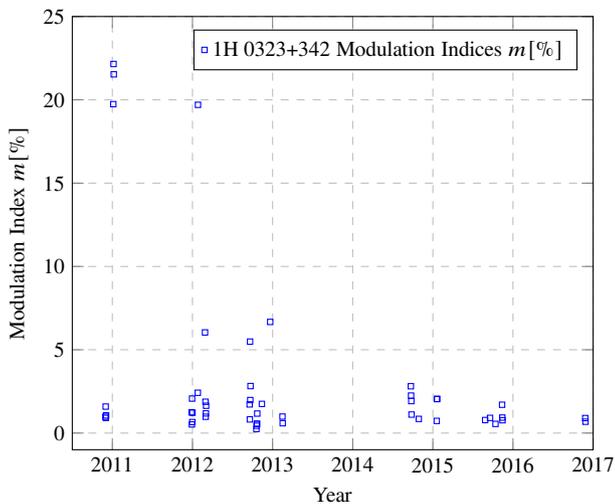

\section{Discussion}
In earlier investigations of microvariations for 1H 0323+342, positive detections of microvariations were reported with amplitudes of the variations ranging from $\approx0.07$ mag  \citep{Paliya2013} to >0.35 mag \citep{Paliya2014}. In the first set of observations, 1H 0323+342 was monitored on four separate nights. On three of those nights, there were significant variations in the seeing (as evidenced from the variations in the FWHM of the point spread function), and thus there is no strong evidence supporting intrinsic variations for 1H 0323+342 on those nights. However, on the night of 01-26-2012, there was significant variability detected with a variation of ~0.07 mag which has a duration of $\approx 2$ hours. The shape of this event is mostly rounded with the peak occurring near ~16.0 U.T. In a second campaign in 2012,\citet{Paliya2014} reported a large microvariability flare on 12-09-2012 when the flux was observed to increase $\approx 0.35$ mag and then approximately 2 hours later it returned to the level prior to the flare. The duration of this flare is $\approx 2$ hours with an overall flat-topped shape to this flaring event. Note that the large variations observed in the FWHM during the night does somewhat reduce the confidence one has in the detail structure of the light curve, but the magnitude of this variation is likely real. \citet{Ojha2019} reported observations of 1H 0323+342 on 02-12-2016 when a rise and decline by $\approx 0.07$ magnitudes were observed with high signal-to-noise observations. The shape of this flare is flat-topped with a rapid rise and fall and an overall duration of $\approx 1.25$ hours. In summary, the duration of these microvariability events in these earlier investigations are typically on the order of a few hours and the range in optical flux is on the order $\approx 0.1$ mag.

In the present investigation, we have, for the first time, identified and positively detected the presence of a low amplitude flux variations with an amplitude of $\approx 0.03-0.04$ mag with a duration of this complete event $\approx 2-3$ hours. It is important to note that the duration of all the detected microvariability events discussed here is quite similar, typically a few hours. This raises the possibility that this represents a basic, fundamental timescale (or physical size) in the source associated with the emitting region.

We know that low amplitude, small timescale events such as the one present in this data can arise from small, localized events in the source region. Events of this form may be due to turbulence that is present in standing shocks in the relativistic jet \citep{Marscher1992}. We also note that the flare has an approximately 3:1
ramp up to ramp down time. This could be due to a density gradient across the standing shock in the jet. These low amplitude variations could also be produced by disturbances other than those in the jet, e.g. hot spots or flares in the accretion disk \citep{Mangalam1993}.

1H 0323+342’s long term optical variations indicate that this AGN underwent a general decline in level of activity over an approximately 3 years interval and then it remained in this lower activity state for the following 4 years. This is seen in both figures \ref{fig:Master} and \ref{fig:MasterModulation}.
However, our limited time-coverage from 2010 to the present does not allow us to determine if this is part of a longer term decline. Near the beginning of our observing time, the AGN exhibited rapid, shorter timescale variations with an amplitude of nearly a magnitude. However, after 3 years, the observed amplitude of the shorter term variations are reduced to around 0.2 magnitudes. During our last year of observation, the short term variations have further declined in amplitude to around 0.1 magnitudes. This long term trend implies observations of this AGN need to be carried out over many years to gain a fuller understanding of the nature of the character of the variations observed for this source and to provide for fuller understanding of both the low frequency and high frequency portion of the power density spectrum, particularly with this light curve containing many multiyear trends.





\section*{Data Availability}

The data for the charts used in this paper may be downloaded from the authors' webpage.

\url{https://www.astro.gsu.edu/~turner/1H0323+342_TotalSet.csv}

\url{https://www.astro.gsu.edu/~turner/1H0323+342_Flare_31inchTelescope.csv}

\url{https://www.astro.gsu.edu/~turner/1H0323+342_Flare_72inchTelescope.csv}

\begin{table}
\caption{Nightly Summary}
\label{tab:NightlySummary}
\begin{tabular}{cccccc}
\hline\hline
{\em Mean JD} & {\em Date} & {$N_{obs}$} & {\em Span(min)} & {\em R Mag} & $m[\%]$\\ \hline
2455530.810   &        11/30/10         &       72     & 539    &     15.922 & 0.98	\\
2455531.806   &         12/1/10         &       72     & 534    &     15.911 & 1.59	\\
2455532.808   &         12/2/10         &       72     & 534    &     15.879 & 0.91	\\
2455533.803   &         12/3/10         &       69     & 534    &     15.934 & 1.06	\\
2455566.648   &          1/5/11         &      144     & 173    &     15.481 & 19.74	\\
2455567.691   &          1/6/11         &      245     & 387    &     15.537 & 22.15 \\
2455568.693   &          1/7/11         &      260     & 387    &     15.538 & 21.53	\\
2455923.689   &        12/28/11         &       96     & 354    &     16.059 & 1.25	\\
2455924.698   &        12/29/11         &       86     & 347    &     16.135 & 0.52	\\
2455925.715   &        12/30/11         &       48     & 325    &     16.129 & 2.07	\\
2455926.735   &        12/31/11         &       42     & 317    &     16.134 & 0.66	\\
2455927.689   &          1/1/12         &       96     & 354    &     16.119 & 1.22	\\
2455951.729   &         1/25/12         &       93     & 187    &     15.558 & 2.42	\\
2455952.633   &         1/26/12         &       93     & 156    &     15.803 & 19.70	\\
2455984.633   &         2/27/12         &      100     & 62     &     16.206 & 6.04	\\
2455986.615   &         2/29/12         &       31     & 23     &     16.310 & 1.88	\\
2455987.655   &          3/1/12         &       19     & 68     &     16.315 & 0.98	\\
2455988.624   &          3/2/12         &       15     & 53     &     16.302 & 1.18	\\
2455989.654   &          3/3/12         &       20     & 94     &     16.302 & 1.64	\\
2456187.904   &         9/17/12         &      541     & 230    &     16.216 & 1.72	\\
2456188.896   &         9/18/12         &       37     & 272    &     16.182 & 0.82	\\
2456189.893   &         9/19/12         &       39     & 278    &     16.187 & 5.49 	\\
2456190.902   &         9/20/12         &       38     & 271    &     16.184 & 1.98	\\
2456191.902   &         9/21/12         &       38     & 269    &     16.170 & 2.82	\\
2456218.802   &        10/18/12         &        6     & 274    &     16.015 & 0.24	\\
2456219.854   &        10/19/12         &       90     & 505    &     16.039 & 0.44	\\
2456220.846   &        10/20/12         &       87     & 497    &     16.026 & 0.56	\\
2456221.847   &        10/21/12         &      151     & 505    &     16.020 & 0.55	\\
2456222.731   &        10/22/12         &       36     & 177    &     16.011 & 1.17	\\
2456243.808   &        11/12/12         &        9     & 520    &     16.007 & 1.75	\\
2456244.781   &        11/13/12         &        3     & 14     &     16.013 & -	\\
2456245.778   &        11/14/12         &        3     & 14     &     15.961 & -	\\
2456246.791   &        11/15/12         &        3     & 14     &     16.030 & -	\\
2456248.791   &        11/17/12         &        3     & 13     &     16.025 & -	\\
2456249.787   &        11/18/12         &        3     & 14     &     16.068 & -	\\
2456281.804   &        12/20/12         &        6     & 294    &     15.888 & 6.67	\\
2456304.831   &         1/12/13         &        3     & 13     &     16.049 & -	\\
2456305.659   &         1/13/13         &        3     & 48     &     16.078 & -	\\
2456310.642   &         1/18/13         &        4     & 53     &     16.030 & -	\\
2456311.634   &         1/19/13         &        4     & 52     &     16.067 & -	\\
2456312.635   &         1/20/13         &        4     & 53     &     16.067 & -	\\
2456336.752   &         2/13/13         &        3     & 13     &     15.785 & -	\\
2456337.652   &         2/14/13         &       12     & 181    &     16.029 & 0.99	\\
2456338.652   &         2/15/13         &       12     & 183    &     16.045 & 0.59	\\
2456339.595   &         2/16/13         &        3     & 13     &     16.070 & -	\\
2456340.595   &         2/17/13         &        3     & 13     &     16.073 & -	\\
2456341.596   &         2/18/13         &        3     & 14     &     16.082 & -	\\
2456342.596   &         2/19/13         &        3     & 13     &     15.791 & -	\\
2456533.884   &         8/29/13         &        3     & 1      &     15.888 & -	\\
2456540.882   &          9/5/13         &        3     & 1      &     15.808 & -	\\
2456567.956   &         10/2/13         &        3     & 7      &     15.988 & -	\\
2456625.793   &        11/29/13         &        3     & 1      &     16.035 & -	\\
2456658.719   &          1/1/14         &        3     & 1      &     15.960 & -	\\
2456663.770   &          1/6/14         &        3     & 3      &     16.011 & -	\\
2456711.621   &         2/23/14         &        3     & 3      &     15.958 & -	\\
2456922.871   &         9/22/14         &       88     & 189    &     15.945 & 2.81	\\
2456923.883   &        9/23/14         &      415      & 376    &     15.923 & 2.25	\\
2456924.876   &         9/24/14         &      381     & 380    &     16.003 & 1.92	\\
2456925.884   &         9/25/14         &      302     & 380    &     16.021 & 1.11	\\
2456958.927   &        10/28/14         &      158     & 524    &     15.892 & 0.85	\\
2457040.716   &         1/18/15         &       20     & 253    &     15.963 & 0.73	\\
2457041.681   &         1/19/15         &       34     & 295    &     15.918 & 2.04	\\
2457042.686   &         1/20/15         &       34     & 300    &     15.944 & 2.05	\\
\hline
\hline
\end{tabular}
\end{table}

\begin{table}
\contcaption{Nightly Summary}
\label{tab:continued}
\begin{tabular}{cccccc}
\hline\hline
{\em Mean JD} & {\em Date} & {$N_{obs}$} & {\em Span(min)} & {\em R Mag} & $m[\%]$\\ \hline
2457261.552   &         8/27/15         &       48     & 111    &     15.816 & 0.78	\\
2457283.841   &         9/18/15         &       36     & 187    &     15.926 & 0.91	\\
2457308.713   &        10/13/15         &       18     & 92     &     15.792 & 0.55	\\
2457338.824   &        11/12/15         &       88     & 588    &     15.755 & 1.70	\\
2457339.820   &        11/13/15         &       88     & 589    &     15.775 & 0.93	\\
2457340.819   &        11/14/15         &       90     & 589    &     15.771 & 0.77	\\
2457372.705   &        12/16/15         &        4     & 16     &     15.947 & -	\\
2457427.609   &          2/9/16         &        2     & 3      &     15.738 & -	\\
2457428.609   &         2/10/16         &        2     & 3      &     15.744 & -	\\
2457429.610   &         2/11/16         &        2     & 3      &     15.735 & -	\\
2457430.610   &         2/12/16         &        2     & 3      &     15.738 & -	\\
2457460.636   &         3/13/16         &        2     & 4      &     15.764 & -	\\
2457462.635   &         3/15/16         &        2     & 4      &     15.738 & -	\\
2457607.885   &          8/7/16         &        3     & 10     &     15.773 & -	\\
2457608.882   &          8/8/16         &        3     & 10     &     15.780 & -	\\
2457609.880   &          8/9/16         &        3     & 10     &     15.773 & -	\\
2457630.821   &         8/30/16         &        3     & 10     &     15.768 & -	\\
2457631.820   &         8/31/16         &        3     & 10     &     15.779 & -	\\
2457717.807   &        11/25/16         &       79     & 544    &     15.812 & 0.90 \\
2457718.807   &        11/26/16         &       79     & 544    &     15.822 & 0.68 \\
2457837.629   &         3/25/17         &        3     & 10     &     15.843 & -	\\
2457839.630   &         3/27/17         &        3     & 10     &     15.777 & -	\\
\hline
\hline
\end{tabular}

\end{table}


\bibliographystyle{mnras}
\bibliography{paper}




\bsp	
\label{lastpage}
\end{document}